\documentclass[doublecol]{epl2} 

\usepackage{dcolumn}
\usepackage{bm}
\usepackage{amsmath,amssymb}

\title{Experimental observation of shear thickening oscillation}
\shorttitle{Experimental observation of shear thickening oscillation} 

\author{Shin-ichiro Nagahiro\inst{1} \and Hiizu Nakanishi\inst{2} \and Namiko Mitarai\inst{3}}
\shortauthor{S. Nagahiro \and H. Nakanishi \and N. Mitarai}

\institute{                    
  \inst{1} Department of Mechanical Engineering, 
Sendai National College of Technology -Natori, Miyagi 981-1239, Japan\\
  \inst{2} Department of Physics, Kyushu University  - 33, Fukuoka 812-8581, Japan\\
  \inst{3} Niels Bohr Institute, University of Copenhagen - Blegdamsvej 17, DK-2100,
Copenhagen \O, Denmark }

\abstract{
We report experimental observation of the shear thickening oscillation,
i.e. the spontaneous macroscopic oscillation in the shear flow of severe
shear thickening fluid.  
Using density-matched starch-water mixture, in the cylindrical shear
flow of a few centimeters flow width, we observed that
well-marked vibrations of the frequency around 20 Hz appear via a
Hopf bifurcation upon increasing externally applied shear stress.
The parameter range and the frequency of the vibration are
consistent with those expected by a simple phenomenological model of
the dilatant fluid.}

\begin{document}

\maketitle

\section{Introduction}

Dense colloids and dense granule-fluid mixtures are
often called dilatant fluids and known to show severe shear thickening,
i.e. its viscosity changes discontinuously by orders of magnitude upon
increasing the shear rate\cite{barnes,fall,jaeger,wagner}.  This is a
source of a variety of unintuitive behaviors of the
media\cite{merkt,ebata,stefan,Lucio}, and might be used for interesting
application like a body armor\cite{lee}, but often causes problems in
industrial situations, such as the damage of mixer motors due to
overloading\cite{barnes}.

Although the severe shear thickening is a property that can be
demonstrated easily with common material like starch and water,
physicists have not reached a general agreement on its microscopic
mechanism.  Originally, the order-disorder transition of the dispersed
particles was proposed\cite{hoffman1, hoffman2, hoffman3, barnes}, but
later the hydrodynamics cluster formation by the lubrication force is
considered to be more consistent with experimental observations and
numerical simulations\cite{bender, Maranzano,brady, melrose}.  More
recently, the {shear thickening} transition is discussed in
connection with the jamming and/or the compaction of
granules\cite{lootens1, brown, bertrand,waitukaitis}.

\begin{figure}
 \centerline{\includegraphics[width=5cm]{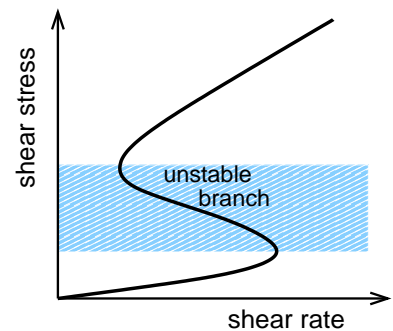}}
\caption{Schematic flow curve for a dilatant fluid that causes a
 discontinuous thickening.  For  the shear stress in the shaded region,
 a uniform shear flow is unstable.
}
\label{FlowCurve}
\end{figure}
{ Recently, the present authors constructed a phenomenological model
that describes macroscopic flowing properties of a shear thickening
fluid\cite{nakanishi1, nakanishi2}.
By analyzing the model, they pointed out that the discontinuous
thickening upon increasing the shear rate signifies that the shear
thickening fluid is actually shear stress thickening, not shear rate
thickening because only the shear stress thickening can produce the
S-shaped flow curve with an unstable branch, that causes the
discontinuous transition experimentally observed (Fig.\ref{FlowCurve}).
Such a flow curve has been suggested also by a semi-microscopic
theory of shear thickening fluid\cite{Holmes-2005}.  }
\begin{figure}
\centerline{\includegraphics[width=6cm]{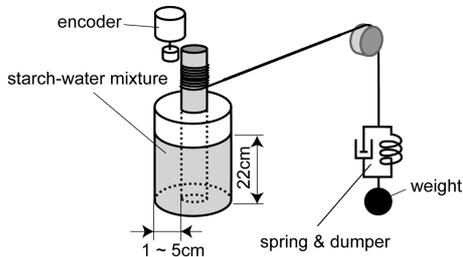}}
\caption{Schematic illustration of the experimental setup consisting of
a cylindrical container, a rotating rod, and a rotary encoder.}
\label{ExpSetup}
\end{figure}
\begin{figure*}
\centerline{\includegraphics[width=16cm]{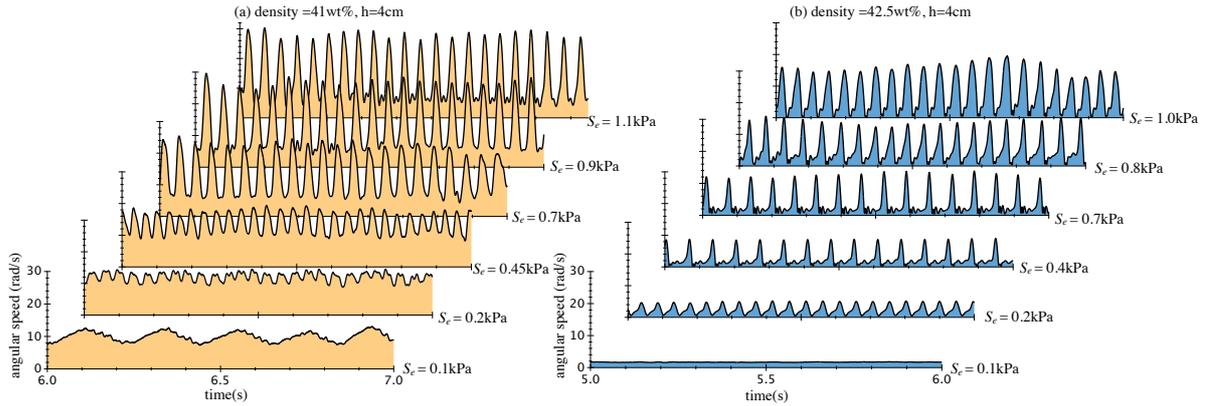}}
\caption{(Color online) Time evolutions of angular speed of the rod for
the flow of  $h=4$ cm thickness with various applied stresses.
Suspension densities are (a) 41 wt\% and (b) 42.5 wt\%.
}
\label{osc_time}
\end{figure*}

{ An interesting consequence of the existence of this unstable
branch is that the flow may show a spontaneous oscillation, i.e.  {\it
the shear thickening oscillation}, when the applied external shear
stress is in the unstable region.  It was predicted using the
phenomenological model\cite{nakanishi1, nakanishi2}, and its
basic mechanism can be understood as follows; A uniform shear flow
{in the unstable branch} can be destabilized by an infinitesimal
disturbance toward the thickened state. Once the thickening starts, the
flow is decelerated, which imposes the inertia stress.  This additional
inertia stress induces further thickening to lead to sudden stopping of
the flow.  However, the fluid starts flowing again as the medium relaxes
because the external stress is not strong enough to maintain the
thickened state.  Such cycles of alternation between the thickened state
and the relaxed state is the shear thickening oscillation.
By estimating the parameters using experimental data for the
cornstarch-water mixture\cite{fall}, the oscillation frequency is
expected to be of the order of 10 Hz when the flow width is several
centimeters.  This oscillation is macroscopic, and is not
due to a discreteness of granules. 

{ The shear thickening oscillation is a salient property predicted
by the phenomenological model and it is actually not difficult to
demonstrate the vibration by just pouring the starch-water mixture out
of a container, but we could not find any published literature which
reports such a clear oscillation except for noisy
fluctuations\cite{laun, lootens1}.
One of the reasons for this might be that a typical size of rheometer
sample for a precision measurement may be too small; The oscillation is
suppressed by the viscosity when the flow width is narrower than
the length scale related to the momentum diffusion length scale
\{associated with the shear rate in the unstable branch.}
%
In this paper, using containers with several centimeters flow width, we
report experimental observations of spontaneous oscillation in the
cylindrical Couette flow of the starch-water mixture.}

\section{Experimental Setup}

Figure \ref{ExpSetup} illustrates our experimental setup.  The fluid
flows in the Taylor-Couette geometry between the outer cylinder and the
coaxial rod at the center.  The cylindrical container is 24 cm tall, but
is filled with the fluid {up to 22 cm deep.}  The diameter of the center
rod is 5 cm;  We use several outer cylinders with different inner
diameters, so that we can have the flow of the width $h =1\sim 5$ cm
thicknesses. The outer cylinder and the rod are both acrylic, and their
surfaces are lined with water-proof sand paper in order to enforce the
no-slip boundary condition.
The center rod rotates under an externally applied constant torque with
the outer cylinder being fixed. The external torque is applied by a
weight through a steel wire wound on the rod; the weight is hung through
a spring and dumper system (Samini Co., Ltd.), and {the mass of the
weight} is in the range of $0.5\sim 10$ kg, which gives the external
stress $S_e =0.14\sim 2.8$ kPa at the surface of the rod. The angular
speed of the center rod $\omega$ is recorded by a rotary encoder
{at the sampling rate $2\times 10^3$ 1/s} (SP-405ZA, Ono Sokki).
As for the fluid, we use the suspension of potatostarch (Hokuren)
density-matched with the aqueous solution of CsCl of the density 1.6
g/cm$^3$.  The upper surface of the fluid is open to the air although
the container is capped with an aluminum plate.

\section{Shear Thickening Oscillation} 

Figure \ref{osc_time} shows the time evolutions of the angular speed of
the rod for the potatostarch suspension of the concentrations 41 wt\%
(a) and 42.5 wt\% (b) with the flow width $h=4$ cm.  The plots show the
oscillations for the various external shear stresses $S_e$ at the
surface of the rod.  For both of the concentrations, clear oscillations
about 20 Hz are observed for the external stress $S_e\gtrsim 0.2$ kPa.
{ As the external stress decreases, the frequency stays almost
constant around 20 Hz, but the amplitude of the oscillation decreases
and {vanishes} at $S_e=0.1$ kPa, thus the transition is consistent
with a Hopf bifurcation as has been expected for the phenomenological
model\cite{nakanishi1, nakanishi2}.  In the case of the thinner fluid of
the 41 wt\% concentration, a slower oscillation around 5 Hz remains at
$S_e=0.1$ kPa (Fig.\ref{osc_time}(a)).  }
For the large external stress, we expect that the rod will be
stuck due to the severe shear thickening, but in the present experiment,
for $S_e\gtrsim 2.0$ kPa, the rod starts slipping suddenly after initial
transient.
{ The shape of the oscillation wave looks rather symmetric and
sinusoidal for the 41 wt\% suspension, while for the 42.5 wt\%
suspension the oscillation is somewhat asymmetric even
at smaller stress $S_e=0.2$ kPa.  The oscillation is 
a sawtooth type that consists of a gradual increase of
the angular speed followed by a sudden drop, and becomes 
asymmetric further for larger $S_e$. }
\begin{figure}
\center{\includegraphics[width=8cm]{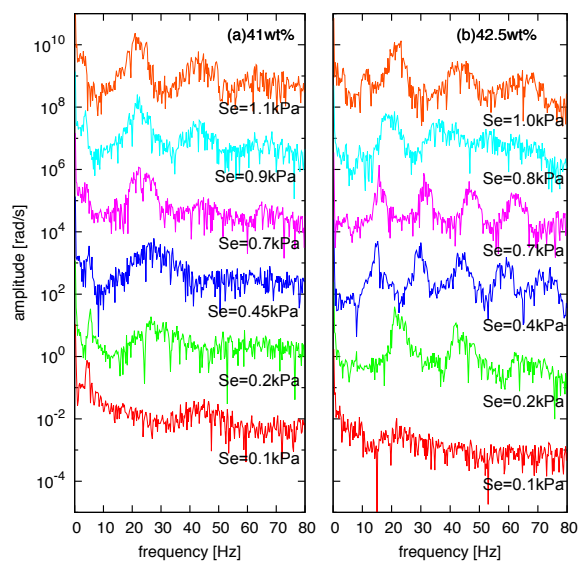}}
\caption{Fourier spectra of the oscillation for the time evolutions of
the oscillation in Fig.\ref{osc_time} for
$S_e =$ 0.1 (bottom), 0.2, 0.45, 0.7, 0.9, and 1.1 kPa (top).  
 The absolute values of the
spectra are plotted in the logarithmic scale, shifting by the factor 100
 for the upper plots to avoid overlapping.
} \label{fourier}
\end{figure}

{ These feature can be seen also in the Fourier space.  Figure
\ref{fourier} shows the Fourier spectra of the time sequences in
Fig.\ref{osc_time}. 
$S_e$, there is a peaks around 20 Hz accompanied by the peak of its
higher harmonics around 40 and 60 Hz.
For the 41 wt\% suspension, the amplitude for the higher harmonics
decreases faster than that for the peak at 20 Hz as $S_e$ decreases,
showing the oscillation becomes closer to the sinusoidal form.  
For the 42.5 wt\% suspension, this tendency is not so clear, and the
basic frequency is not stable but changes between 15 Hz $\sim$ 25 Hz.
%
}

\begin{figure}

\center{\includegraphics[width=8cm]{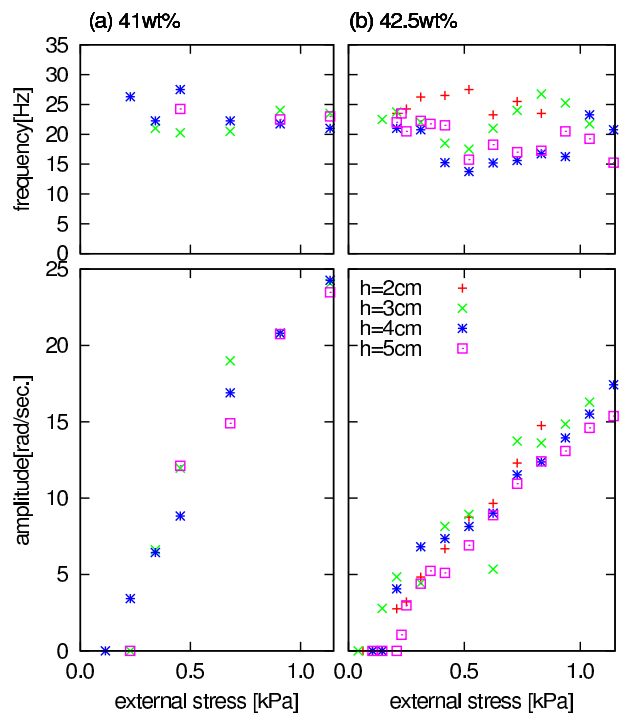}}
\caption{
Stress dependences of the frequency (upper) and the amplitude (lower)
of the oscillation around 20 Hz for various flow widths $h$.
}
\label{amp_freq}
\end{figure}

{ The amplitude and the frequency of the peaks around 20 Hz are
plotted as a function of $S_e$ for some flow widths $h$ in
Fig.\ref{amp_freq}(a) and (b), respectively.  There exists a threshold
of $S_e$ around 0.1 kPa toward which the amplitude vanishes continuously
as $S_e$ decreases. On the other hand, the frequency stays almost
constant near the threshold, suggesting the bifurcation is the Hoph
type.
}

\begin{figure}
\centerline{\includegraphics[width=8cm]{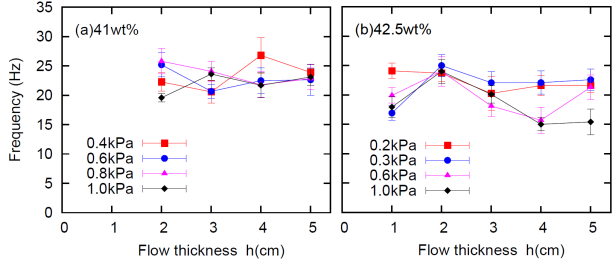}}
\caption{(Color online) The frequency of the shear thickening
oscillation for (a) 41 wt\% and (b) 42.5 wt\% potatostarch suspension
as a function of flow thickness $h$.  The plots are evaluated from the
average interval of neighboring peaks, and the errorbars represent
standard deviation.
}  \label{freq_h}
\end{figure}

{It should be notable that both the amplitude and the frequency}
barely depend on the flow width $h$ within the parameter range studied
in the present experiments.
Figure \ref{freq_h} shows the oscillation frequency as a function of the
flow width $h$ for various external stresses $S_e$.  One can see that
the frequencies are around 20 Hz for all the cases, and we could not
find any systematic dependence on neither $h$ nor $S_e$.  For $S_e$
higher than 1.0 kPa, the center rod starts slipping and we could not
obtain clean data, but we find no clear sign of a systematic change in
the frequency up to $S_e = 2.3$ kPa.

\begin{figure}
\centerline{
\includegraphics[width=8cm]{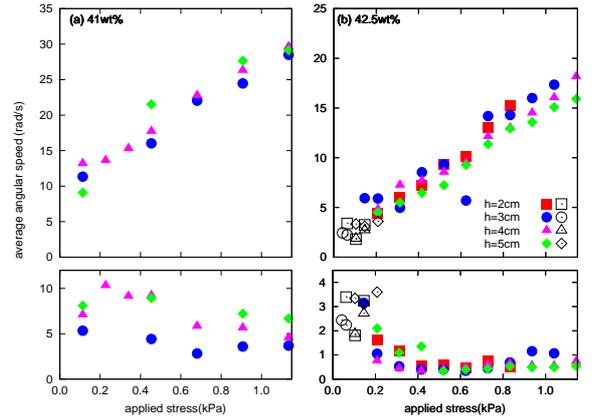}}
\caption{(Color online) The maximum and minimum angular speed of the
center rod during the oscillations as a function of the applied stress
$S_e$ for (a) 41 wt\% and (b) 42.5 wt \% potatostarch suspensions.
Each data point represents an average over a single run of experiment,
which contains typically $10^2$ cycles of the oscillation.  The open
symbols in (b) represent the angular speed of the non-oscillating
flows for small $S_e$.  } 
\label{omega_min-max-stress}
\end{figure}

The maximum and minimum angular speed of the rod, $\omega_{\rm max}$ and
$\omega_{\rm min}$, during a single cycle of oscillation are plotted
against $S_e$ in Fig.\ref{omega_min-max-stress} for some values of the
flow width $h$ for both of the concentrations. Each data point
represents an average over a single run of the experiment, which
contains typically $10^2$ cycles of the oscillation.
%
%
Neither $\omega_{\rm max}$ nor $\omega_{\rm min}$ depend on the
flow width $h$ in any appreciable way.  
As for the maximum angular speed $\omega_{\rm max}$,
the plots show almost linear dependence on  the external
shear stress $S_e$:
\begin{equation}
\omega_{\rm max}\approx a S_e + \omega_0
\end{equation}
with $a\approx 11~{\rm rad/(kPa\cdot s)}$ and 
$\omega_0 \approx 10~{\rm rad/s}$ for the 41 wt\% 
suspension (Fig.\ref{omega_min-max-stress}(a)), and 
$a\approx 15~{\rm rad/(kPa\cdot s)}$ and 
$\omega_0 \approx 0$ for the 42.5 wt\% suspension
(Fig.\ref{omega_min-max-stress}(b)).
On the other hand, the minimum angular speed $\omega_{\rm min}$ stays
roughly constant for $S_e\gtrsim 0.5$ kPa after the initial decrease for
smaller $S_e$.
In the case of the 42.5 wt\% suspension, the oscillation disappears
around $S_e\lesssim 0.2$ kPa, for which cases the terminal angular
speeds are plotted with open marks in Fig.4(b).

The linear dependence of $\omega_{\rm max}$ on $S_e$ leads us to
introduce  an average viscosity
$\eta_{\rm relax}$ for the
flow during the relaxed state.
If we 
assume the steady cylindrical Couette flow with a constant angular
velocity $\omega$, the viscosity would be given by
\begin{equation}
\eta = {S_e\over 2\omega}\left[1-\left({R_1\over R_2}\right)^2\right],
\label{eta}
\end{equation}
where $R_1=2.5$ cm is the radius of the center rod and $R_2=R_1+h$ is
the radius of the outer cylinder.
Substituting $\omega_{\rm max}$ of Eq.(1) into $\omega$ in Eq.(2),
{we have  $\eta_{\rm relax}\sim 30 {\rm Pa\cdot s}$ }
for the 42.5 wt\% suspension; This is consistent with the previous
viscosity measurement before the thickening transition\cite{fall}.
For the 41 wt\% suspension, Fig.\ref{omega_min-max-stress}(a) shows a
substantial non-zero extrapolation $\omega_0$ at $S_e=0$. This may be
due to the inertia effect in the oscillatory state, in which case our
steady state formula, Eq.(\ref{eta}), is not valid for this case.

In the thickened state, $\omega_{\rm min}$ is nearly constant around
$0.5$ rad/s for $0.5\lesssim S_e\lesssim 1.5$ kPa for the 42.5wt\%
suspension.  The estimated values of the viscosity in the thickened
state are of the order of $10^3$ Pa$\cdot$s, which seems to be somewhat
smaller than the value observed in Ref.\cite{fall}.  This may suggest
that the whole system is not thickened uniformly, but the thickening
region is localized as we will discuss later.


\section{Discussions}

We experimentally observed the spontaneous oscillation of
macroscopic shear flow in the shear thickening fluid.  The shear flow
starts oscillating with a finite frequency through a Hopf
bifurcation upon increasing the externally applied shear stress.  The
oscillation is asymmetric in the sawtooth shape, i.e. a gradual increase
followed by a sudden deceleration, but seems to become more symmetric
for the stress near the threshold.  The observed frequency of the
oscillation is about 20 Hz, which is roughly in the range expected by
the model for the flow width of several centimeters.  These
features are consistent with those of the shear thickening oscillation
predicted by the phenomenological model\cite{nakanishi2}.

On the other hand, there are some other features that seem difficult to
interpret in terms of the simple shear flow oscillation {as has been
analyzed for the model.}  The oscillation frequency does not depend
significantly neither on the external stress $S_e$ nor on the flow width
$h$,
whereas the simulations of cylindrically symmetric oscillation show that
the frequency tends to become slightly higher as $S_e$ increases, and
that the frequency decreases significantly as the flow width $h$
increases.
It is also tricky to interpret the behavior of the maximum $\omega_{\rm
max}$ and minimum angular speed $\omega_{\rm min}$ of the oscillation:
$\omega_{\rm max}$ is proportional to $S_e$ while $\omega_{\rm min}$
stays almost constant, and neither of them depends on the flow width
$h$.  We find that these features are not easy to reproduce by the model
as long as we assume that the flow is cylindrically symmetric.

{ One possibility is that thickening does not occur uniformly, thus
these experimentally observed parameter dependences cannot be
interpreted using  the cylindrically symmetric flow. 
At the moment, we cannot determine the symmetry of the flow in our
experiments because we cannot observe the thickening region in the
cylinder.  {We performed numerical simulations in two dimensions on
the phenomenological model, and we found that the cylindrical Poiseuille
flow is destabilized for larger shear stress and a localized thickened
band region appears diagonally to support most of the shear
stress\cite{private}.  We have not performed simulations in three
dimensions yet, but it is natural to expect such localized thickened
bands to appear also in three dimensions because there is a natural
tendency for the shear stress thickening fluid to form a thickened band;
Once a thickened region appears, it supports larger stress, which
thickens the region further.  }}
%

The oscillation we report here is not subtle but quite evident
under an appropriate condition, yet we could not find any previous
reports on this phenomenon.  
A couple of remarks on unstable fluctuations\cite{laun, lootens1},
which might be related to the present oscillation, are
the only things that we could find in the published literature.  
We suspect that this
is because most of the previous measurements on the shear thickening
fluid have been done on samples with the flow width of the order of
millimeters, while the macroscopic oscillation is suppressed in the
shear flow narrower than the length scale related to the momentum
diffusion length scale, which is estimated to be of the order of
centimeters in the present material\cite{nakanishi1, nakanishi2}.

{It should be mentioned that} there is a report on an oscillatory shear flow
of a lyotropic lamellar phase\cite{Wunenburger}.  The phenomenon is
analogous in the sense that the internal structure of the fluid changes
during a period of oscillation, and the change in the viscosity
associated with the structural change seems to cause the oscillation,
thus a similar phenomenological model may be used to describe the
behavior.  On the other hand, the differences from the present
oscillation are that the material shows shear thinning and the period of
oscillation is of the order of a thousand seconds, thus no fluid
dynamical inertia effect should be involved in the oscillation mechanism
unlike in the case of the shear thickening oscillation. 

\section{summary}
In summary, using the starch-water mixture, we observed the shear
thickening oscillation with the frequency around 20 Hz in the shear flow
of the Taylor-Couette geometry with several centimeters flow width.
{The oscillation starts through a Hopf bifurcation with the
frequency on increasing the external shear stress.  The frequency and
the amplitude of the oscillation barely depend on the flow width for
the range 1$\sim$ 5 cm of the present experiments.}

\acknowledgments
We thank T. Kato, T. Sugawara, and Y. Fukuda for technical assistance.
This work is supported by KAKENHI grant number 21540418 (H.N.) and
24760145 (S. N.), and the Danish Council for Independent Research,
Natural Sciences (FNU) (N.M.).

\bibliography{sto}

\end{document}